\documentclass[aps,pra,twocolumn,showpacs,superscriptaddress,floatfix,10pt]{revtex4-1}
\usepackage{framed}
\usepackage{graphicx}
\usepackage{amsmath}
\usepackage{epsfig}
\usepackage{helvet}
\usepackage{amssymb}
\usepackage{framed}
\usepackage{bbold}
\usepackage{subfigure}

\newcommand{\ket}[1]{\mbox{$| #1 \rangle$}}

\allowdisplaybreaks[3]

\def\appendix{Appendix }

\begin{document}
\title{TensorNetwork on TensorFlow:\\
A Spin Chain Application Using Tree Tensor Networks}

\author{Ashley Milsted}
\affiliation{Perimeter Institute for Theoretical Physics, Waterloo, ON, Canada}

\author{Martin Ganahl}
\affiliation{Perimeter Institute for Theoretical Physics, Waterloo, ON, Canada}

\author{Stefan Leichenauer}
\author{Jack Hidary}
\affiliation{Alphabet (Google) X, Mountain View, CA 94043, USA}

\author{Guifre Vidal}
\affiliation{Perimeter Institute for Theoretical Physics, Waterloo, ON, Canada}
\affiliation{Alphabet (Google) X, Mountain View, CA 94043, USA}

\begin{abstract}
TensorNetwork is an open source library \cite{download} for implementing tensor network algorithms in TensorFlow. 
We describe a tree tensor network (TTN) algorithm for approximating the ground state of either a periodic quantum spin chain (1D) or a lattice model on a thin torus (2D), and implement the algorithm using TensorNetwork. We use a standard energy minimization procedure over a TTN ansatz with bond dimension $\chi$, with a computational cost that scales as $O(\chi^4)$. Using bond dimension $\chi \in [32,256]$ we compare the use of CPUs with GPUs and observe significant computational speed-ups, up to a factor of $100$, using a GPU and the TensorNetwork library.
\end{abstract}

\maketitle

%\tableofcontents
%TO DO:
%\begin{itemize}
%\item Add correct references for TensorFlow, TensorNetwork, TTN code 
%\end{itemize}

\section{Introduction}

\begin{figure}
  \includegraphics[width=8cm]{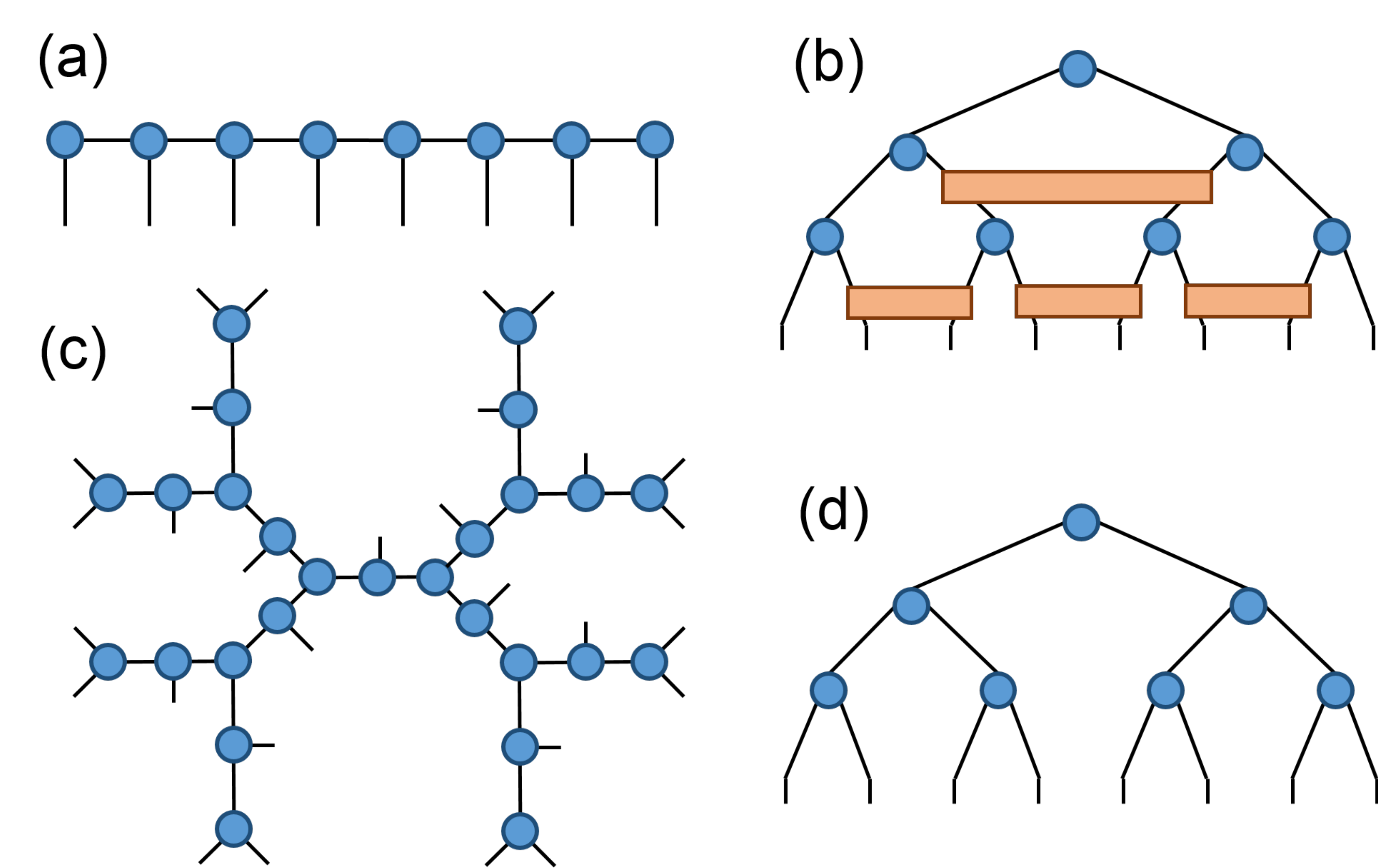}
\caption{
(a) \textit{Matrix product state} (MPS) for a many-body wave-function on $L=8$ sites.
(b) \textit{Multi-scale entanglement renormalization ansatz} (MERA) also for the state of a lattice system made of $L=8$ sites.
(c) Example of a \textit{tree tensor network} (TTN), where the network of tensors is organized according to a tree structure. Notice the absence of closed loops, as in the MPS and in contrast to the MERA.
(d) The specific TTN considered in this paper: a regular binary tree. Like the MPS, it is loop-free. Like the MERA, it is organized in layers of tensors corresponding to different length scale.
\label{fig:TTN} 
}
\end{figure}

Tensor networks are sparse data structures originally developed to efficiently simulate complex quantum systems in condensed matter \cite{Fannes, White, Vidal, Perez-Garcia, MERA, MERA2, MERAalgorithms, Shi, Tagliacozzo, Murg, PEPS1, PEPS2, PEPS3, rev1, rev2, rev3, rev4, rev5}. In recent years, highly successful tensor networks such as the \textit{matrix product state} (MPS) \cite{Fannes, White, Vidal, Perez-Garcia} and the \textit{multi-scale entanglement renormalization ansatz} (MERA) \cite{MERA, MERA2, MERAalgorithms} (see Figs. \ref{fig:TTN}(a)-(b)) have found a much wider range of applications, including quantum chemistry \cite{QC1, QC2, QC3, QC4}, statistical mechanics \cite{CTMRG, TRG, TEFRG, TNR}, machine learning \cite{ML1, ML2, ML3, ML4, ML5}, quantum fields \cite{cMPS, cMERA}, and even quantum gravity and cosmology \cite{Swingle, dS1, dS2, dS3, MERAgeometry}.
 
TensorFlow \cite{TensorFlow} is a free, open source software library for dataflow and differentiable programming, developed by the Google Brain team, that can be used for a range of tasks including machine learning applications such as neural networks. Recently, the open source library TensorNetwork \cite{download} has been released to allow running tensor network algorithms on TensorFlow. 

This paper is one of a series of papers that aim to illustrate, with examples of tensor network algorithms, the use of TensorNetwork in actual computations. Specifically, here we describe an algorithm for approximating the ground state of a periodic quantum spin chain or thin torus with a tree tensor network (TTN) \cite{Shi, Tagliacozzo, Murg}, which is a tensor network where the tensors are connected according to a tree structure. We use a standard energy minimization algorithm, whose code can be downloaded here \cite{download}. Companion papers will present other algorithms, including MPS and MERA algorithms.

The specific TTN for 1D quantum systems that we consider here, represented in Fig. \ref{fig:TTN}(d), lies in some sense between an MPS and the MERA, depicted in Figs. \ref{fig:TTN}(a) and  \ref{fig:TTN}(b), respectively \cite{Comparison}. Like the MPS, a TTN has no closed loops, and this allows for an optimal compression of each bond index of the tensor network (and thus also of each tensor) using the Schmidt decomposition. Like the MERA, however, the TTN in Fig. \ref{fig:TTN}(d) organizes the tensors in an additional (vertical) dimension corresponding to scale. One can think of this TTN as a simplified version of the MERA in which a subset of tensors, called disentanglers, have been removed. The advantage of a TTN over MERA is that the absence of disentanglers makes it conceptually simpler. TTN algorithms are also more easily generalized from 1D to 2D systems than MPS or MERA algorithms. These properties make the TTN a good starting point to demonstrate TensorNetwork.

\section{Tree Tensor Network for ground states of lattice models}

\begin{figure}
  \includegraphics[width=8cm]{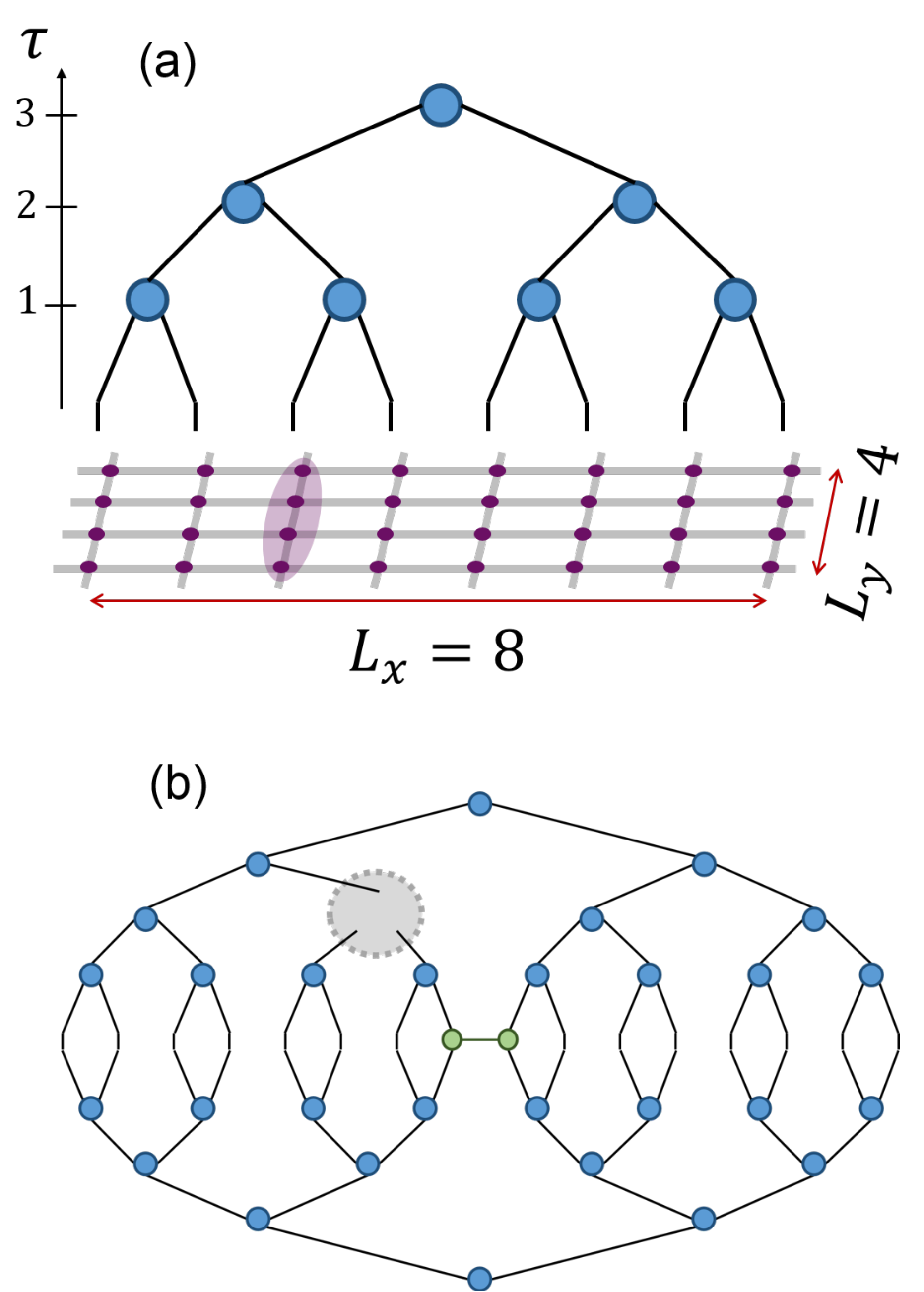}
\caption{
(a) TTN variational ansatz for the ground state of a square lattice of $L_x \times L_y = 8 \times 4 = 32$ quantum spins with toric boundary conditions. Notice that a single dangling leg of the TTN is an index of dimension $d^{L_y} = 2^4 = 16$ that labels an orthonormal basis in the $16$-dimensional Hilbert space $(\mathbb{C}^{2})^{\otimes L_y}$ of $L_y=4$ quantum spins.
(b) Example of diagram needed in order to compute the so-called \textit{environment} $E$ for an isometry $w$, which would be placed in the empty location indicated by a grey shadow. This example corresponds to a TTN for $L_x=16$ and for a Hamiltonian term $H_{m,m+1}$ (in green) connecting effective sites $m=8$ and $m+1=9$. The bond index connecting the two green circles has dimension $L_y$ (for the Ising Hamiltonian) instead of $\chi$.
\label{fig:TreeOnTorus} 
}
\end{figure}

\subsection{Thin torus}

We use the TTN as an approximation or \textit{variational ansatz} for the ground state of a periodic square lattice made of $L_x \times L_y$ quantum spins. Here $L_x$ and $L_y$ denote the length (in units of the lattice spacing) of the lattice in the $x$ and $y$ directions. We consider lattices corresponding to a thin torus, with $L_x \gg L_y$, which for $L_y=1$ turns into a periodic quantum spin chain. We label lattice sites with a pair of integers $(m,n)$, with $1 \leq m \leq L_x$ and $1 \leq n \leq L_y$, and assign a complex vector space $\mathbb{C}^d$ of dimension $d=2$, representing a spin $1/2$ degree of freedom, to each lattice site. 

As a concrete example, we consider the Ising model with transverse magnetic field, with Hamiltonian
\begin{eqnarray} \label{eq:H2D}
H &=& \sum_{m=1}^{L_x} \sum_{n=1}^{L_y}\left( -X_{(m,n)} X_{(m+1,n)}  \right. ~~~~~~~~~~~~~~~~~~~\\
 &&~~~~~~~~~~\left. - X_{(m,n)}X_{(m,n+1)} + h  Z_{(m,n)}\right),
\end{eqnarray}
where $X = \left(\begin{array}{cc} 0 & 1 \\ 1 & 0 \end{array} \right)$, $Z = \left(\begin{array}{cc} 1 & 0 \\ 0 & -1 \end{array} \right)$ are Pauli matrices and $h \in \mathbb{R}$ denotes the strength of a transverse magnetic field. For concreteness, we choose $h=2.9$, for which find a ground state that is entangled over many length scales, see Fig. \ref{fig:EntEntropy}. 
%Notice that this value of the magnetic field is not far from the critical value $h_C\approx 3.04$ of the transverse magnetic field for the 2D quantum Ising model on the plane, that is for $L_x,L_y \rightarrow \infty$.

\subsection{Effective quantum spin chain}

As Fig. \ref{fig:TreeOnTorus}(a) shows for $L_x=8$ and $L_y =4$, each open index at the bottom of the TTN is assigned a Hilbert space $(\mathbb{C}^{d})^{\otimes L_y}$ of dimension $d^{L_y} = 2^{L_y}$ corresponding to the $L_y$ sites $(m,1)$, $(m,2)$, $\cdots$, $(m,L_y)$ at fixed value $m$ of the $x$ direction. Therefore from the perspective of the TTN, the lattice model is effectively a quantum spin chain with $L_x$ sites $m=1,2, \cdots L_x$, with each effective site corresponding to a complex vector space $\mathbb{C}^{d^{L_y}}$ of dimension $d^{L_y}$ and with Hamiltonian 
\begin{equation} \label{eq:H1D}
H = \sum_{m=1}^{L_x} H_{m,m+1},
\end{equation}
where $H_{m,m+1}$ collects all the Hamiltonian contributions connecting effective sites $m$ and $m+1$. For instance, for $L_y=3$ and $m=8$ we have
\begin{eqnarray}
H_{8,9} &=& -X_{(8,1)}X_{(9,1)} - X_{(8,2)}X_{(9,2)} - X_{(8,2)}X_{(9,2)} \label{eq:1}\\
&+& \frac{1}{2}\left\{- X_{(8,1)}X_{(8,2)} - X_{(8,2)}X_{(8,3)} - X_{(8,3)}X_{(8,1)} \right.~~~ \label{eq:2}\\
&&~~~~~~~~~~~~~~+ \left. h \left(Z_{(8,1)} + Z_{(8,2)}+ Z_{(8,3)}\right) \right\}\label{eq:3}\\
&+& \frac{1}{2}\left\{- X_{(9,1)}X_{(9,2)} - X_{(9,2)}X_{(8,3)} - X_{(9,3)}X_{(9,1)} \right.~~~\label{eq:4}\\
&&~~~~~~~~~~~~~~+ \left. h \left(Z_{(9,1)} + Z_{(9,2)}+ Z_{(9,3)}\right) \right\}, \label{eq:5}
\end{eqnarray}
where \eqref{eq:1} collects couplings between pairs of spins, with one spin in column $m=8$ and the other spin in column $m+1=9$; \eqref{eq:2} and  \eqref{eq:3} correspond to interactions and magnetic fields of spins within column $m=8$; finally  \eqref{eq:4} and \eqref{eq:5} correspond to interactions and magnetic fields within column $m+1=9$. The factor $1/2$ is included to avoid double counting in Eq. \eqref{eq:H1D}.

\subsection{The tensor network}

The TTN represents a pure state $\ket{\Psi} \in (\mathbb{C}^2)^{\otimes L_x \times L_y}$ and is made of isometric tensors $w$, or \textit{isometries}, which are rank-3 tensors of size $\chi \times \chi \times \chi$ (here we assume, for simplicity in the explanation, that all the bond dimensions in the TTN are the same and given by $\chi$) and components $w^{\alpha}_{\beta \gamma}$ that fulfil
\begin{equation} \label{eq:constraint}
\sum_{\gamma,\beta=1}^{\chi} w^{\alpha}_{\beta \gamma} \left(w^{\alpha'}_{\beta \gamma}\right)^* = \delta_{\alpha \alpha'} ~~~~~\mbox{(isometric constraint)}.~~
\end{equation}
There is also a rank-2 tensor $v$ at the top of the TTN, which is normalized to 1,
\begin{equation} \label{eq:normalization}
\sum _{\alpha,\beta=1}^{\chi} |v_{\alpha\beta}|^2 = 1 ~~~~~\mbox{(normalization)},
\end{equation}
and fixes to 1 the normalization of the wavefunction $\ket{\Psi} \in (\mathbb{C}^2)^{\otimes L_x \times L_y}$. The isometric constraint \eqref{eq:constraint} and the normalization \eqref{eq:normalization} are represented diagrammatically in Fig. \ref{fig:TensorManipulations}(a). 

We label the isometries in the TTN as $w^{(\tau,m)}$ where $\tau = 0,\cdots, \log_2(L_x) - 2$ labels the scale direction, with $\tau=0$ at the bottom of the TTN and $\tau=\log_2(L_x) -2$ at the top, whereas $m=1,\cdots, L_x/2^\tau$ labels the position within layer $\tau$.
There are $L_x/2$ isometries $w^{(1,1)}$, $\cdots$, $w^{(1,L_x/2)}$ at the lowest layer of the TTN, $L_x/4$ isometries $w^{(2,1)}$, $\cdots$, $w^{(2,L_x/4)}$ at the second lowest layer of the TTN, etc. The total number of isometries is thus $L_x/2 + L_x/4 + \cdots = L_x - 2$. For instance, in the example of Fig. \ref{fig:TreeOnTorus}(a), there are $L_x/2 = 8/2 = 4$ isometries $w^{(1,1)}$, $w^{(1,2)}$, $w^{(1,3)}$, and $w^{(1,4)}$ in the lowest layer of the TTN, and $L_x/4 = 8/4 =2$ isometries $w^{(2,1)}$ and $w^{(2,2)}$ in the second lowest layer. Finally, there is also the rank-2 tensor $v$ at the top of the TTN.

\begin{figure}
  \includegraphics[width=8cm]{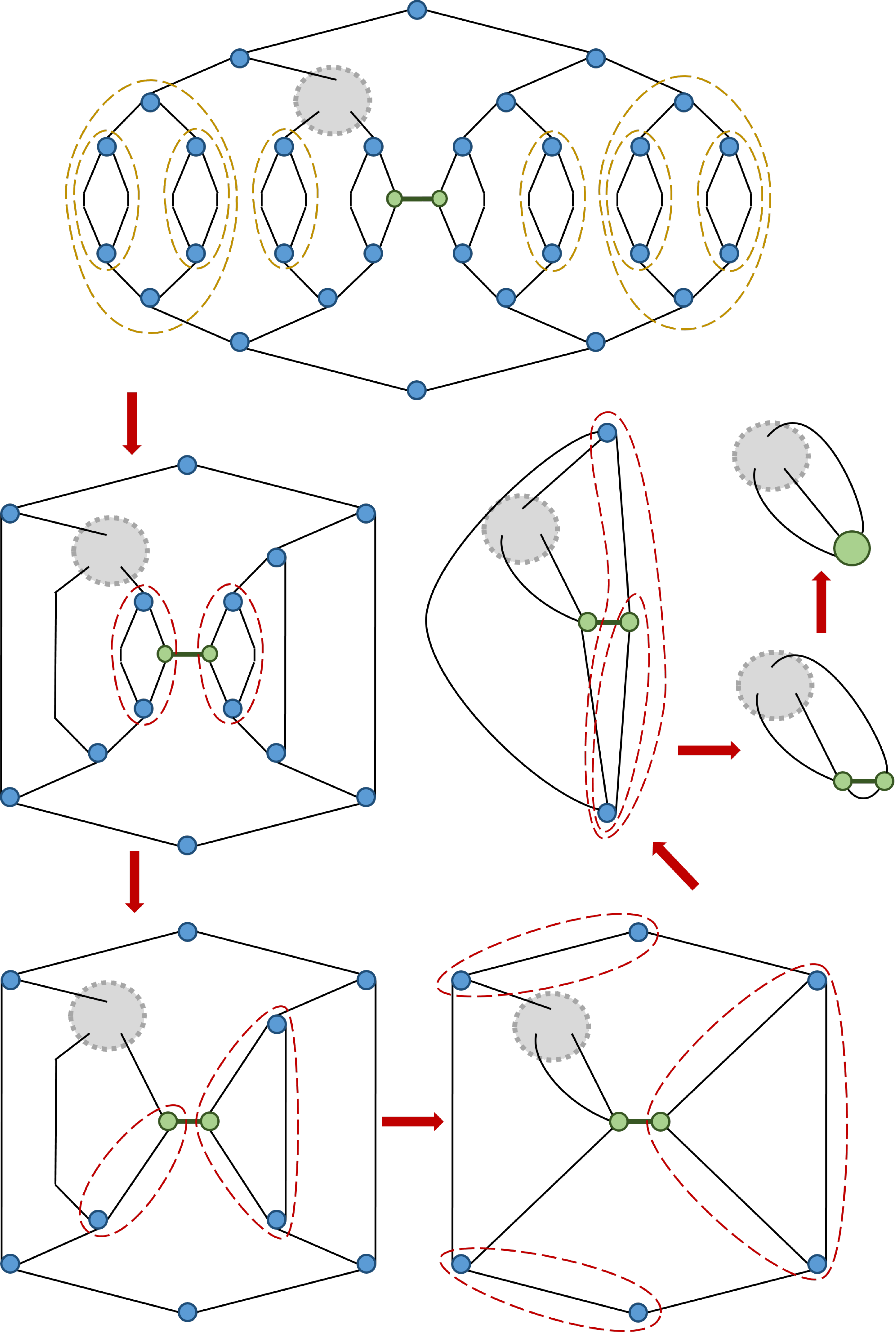}
\caption{
Contraction of the tensor network in Fig. \ref{fig:TreeOnTorus}(b),  corresponding to one contribution to the environment $E$ for a given isometry $w$. In a first step, we use the isometric constraints of isometries in Fig. \ref{fig:TensorManipulations}(a) to eliminate pairs $w,w^{\dagger}$ and thus simplify the network (no actual tensor-tensor contractions need to be computed). The rest of steps can be ultimately decomposed into tensor-tensor contractions at cost $O(\chi^4)$, see Fig. \ref{fig:TensorManipulations}(b). 
\label{fig:Environment} 
}
\end{figure}

\begin{figure}
  \includegraphics[width=6cm]{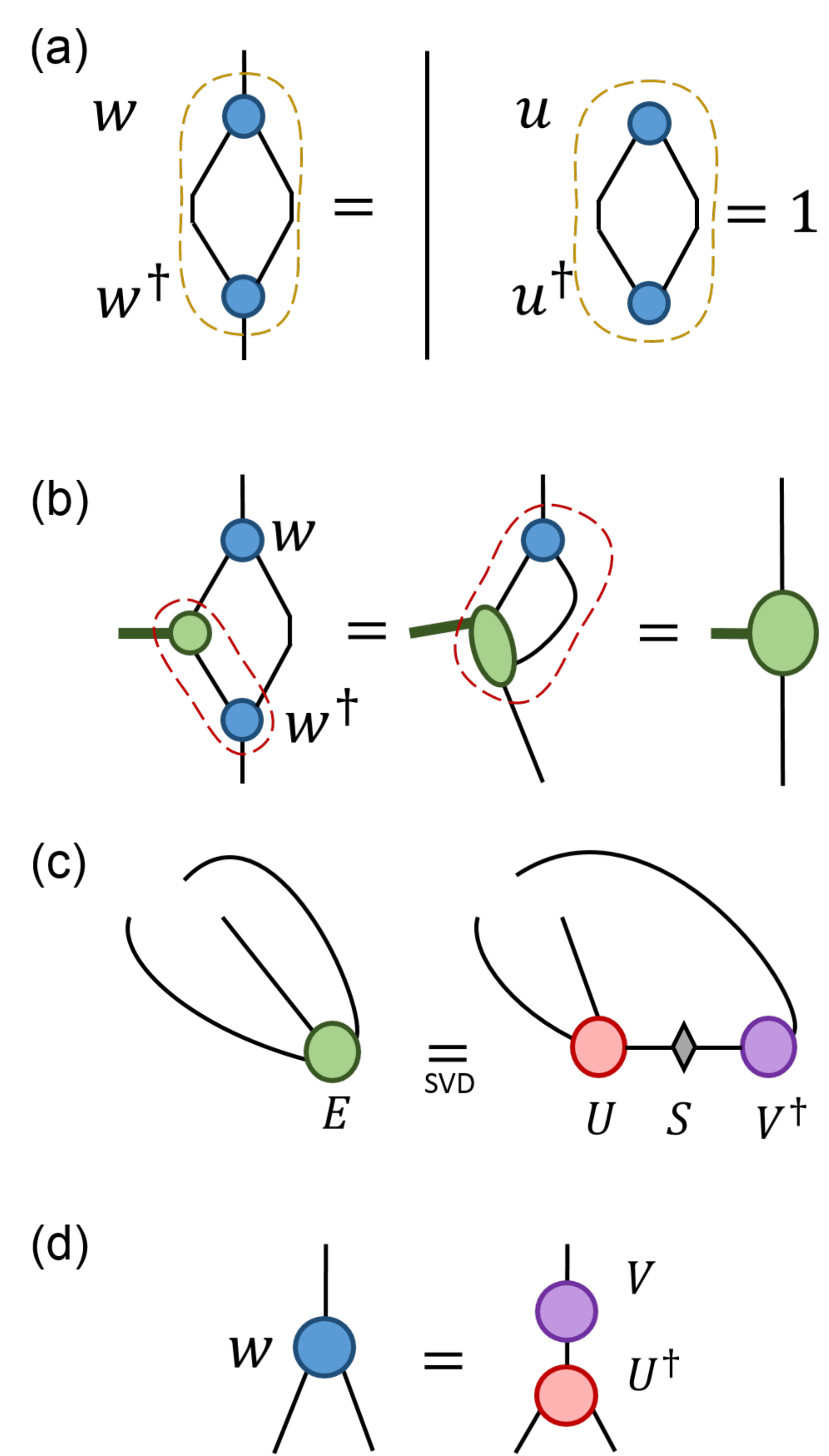}
\caption{
(a) Isometric constraint of an isometry $w$ and normalization of the top tensor $v$, see Eqs. \eqref{eq:constraint} and \eqref{eq:normalization}.
(b) Example of tensor-tensor contractions needed in Fig. \ref{fig:Environment}, at computational cost $O(\chi^4)$.
(c) SVD decomposition of the environment $E$ of an isometry $w$.
(d) Updated isometry $w$ in terms of the tensors $V$ and $U$ that appear in the SVD of the environment $E$.
\label{fig:TensorManipulations} 
}
\end{figure}

\section{Algorithm}

The TTN is optimized using a standard energy minimization algorithm, as described e.g. in section IV of Ref. \cite{Tagliacozzo}. The energy minimization algorithm proceeds by iteratively updating each isometry in the TTN, as outlined below. We exploit translation invariance of $H$ to set all the isometries in a given layer of the TTN to be the same, that is $w^{(\tau,1)} = w^{(\tau,2)} = \cdots$, and therefore the iterative update only progresses through scale, as parametrized by the integer $\tau$ (and not through space, parametrized by the integer $m$).

\subsection{Environment $E$ of an isometry $w$}

In order to update an isometry $w$ of the TTN, we first need to compute its environment $E$. Like the isometry $w$, the environment $E$ is a rank-3 tensor of dimensions $\chi \times \chi \times \chi$. It is defined as a sum of a number of contributions, coming from different Hamiltonian terms $H_{m,m+1}$. An example of such contributions is represented in Fig. \ref{fig:TreeOnTorus}(b). 

The computation of the environment $E$ is achieved by contracting the tensor networks for all relevant contributions. Fig. \ref{fig:Environment} shows a sequence of diagrams corresponding to the contraction of the tensor network in Fig. \ref{fig:TreeOnTorus}(b). Such a tensor network can be contracted using the \textit{ncon} function in TensorNetwork. In practice, contracting the whole network is reduced to a sequence of tensor-tensor contractions. Some of these contractions are trivial due to the isometric constraint and do not need to be implemented, whereas some contractions must be explicitly performed, see Fig. \ref{fig:TensorManipulations}(a)-(b). The latter correspond, possibly after flattening the indices of the tensors, to matrix-matrix multiplications, with computational cost of at most $O(\chi^4)$ per multiplication.  

\subsection{Updated isometry}

Once the environment $E$ for an isometry $w$ has been computed, we flatten the rank-3 tensor $E$ into a $\chi^2 \times \chi$ matrix (which we also refer to as $E$) and apply a singular value decomposition to it, $E = U S V^{\dagger}$. Then we build the $\chi\times \chi^{2}$ matrix $w = V U^{\dagger}$, which we turn into the updated rank-3 isometry $w$ by splitting its second index into two, see Fig \ref{fig:TensorManipulations}(c)-(d). The top tensor $v$ is updated similarly.

\section{Benchmark results}

We consider a 2D lattice made of $L_x \times L_y = 128 \times 5$ quantum spins or, equivalently, a 1D lattice made of $L_x = 128$ effective spins, each of dimension $2^5=32$. We choose the value $h=2.9$, which is seen to lead to a scaling of ground state entanglement entropy compatible with being near a quantum critical point. This value is slightly below $\approx 3.04$, which  corresponds to the critical point in a fully 2D lattice, that is for $L_x,L_y \rightarrow \infty$. We approximate the ground state using a TTN for increasing values of $\chi$ in the range $32 \leq \chi \leq 256$. For each value of $\chi$ we minimize the expectation value of the energy per site by iterating the isometry update scheme outlined above, until the energy per site changes by less than $10^{-10}$ after a whole sweep of updates.

\subsection{Ground state energy}

\begin{figure}
  \includegraphics[width=8cm]{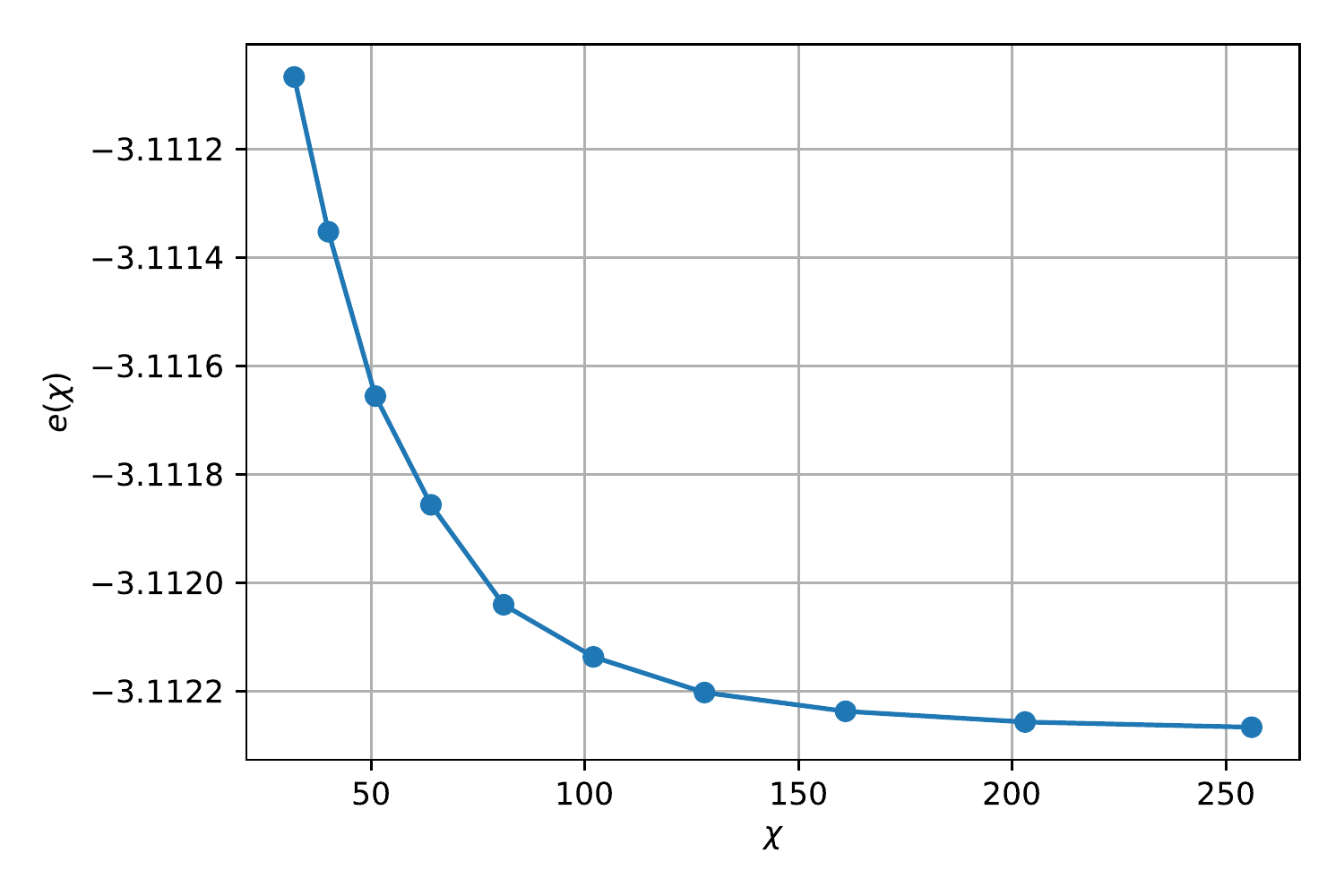}
\caption{
Variational ground state energy per site $e(\chi)$ as a function of the bond dimension $\chi$. 
\label{fig:Energy} 
}
\end{figure}

\begin{figure}
  \includegraphics[width=8cm]{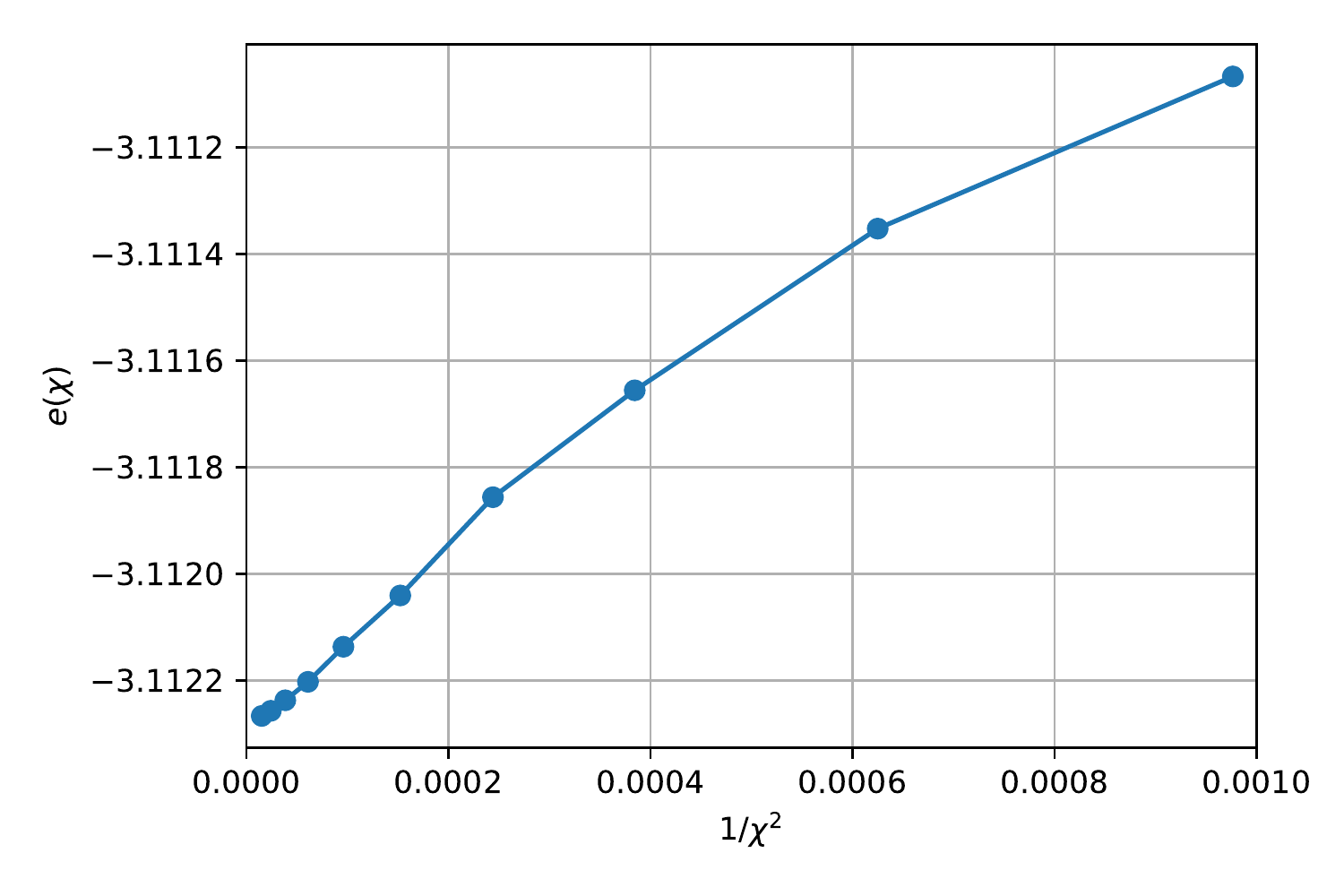}
\caption{
Variational ground state energy per site $e(\chi)$ as a function of $1/\chi^2$. At large bond dimension $\chi$, the energy seems to be scaling to some $\chi = \infty$ limit value $e(\infty)\approx -3.11229$ as $1/\chi^2$. 
\label{fig:Energy2} 
}
\end{figure}

Fig. \ref{fig:Energy} shows the converged value of the ground state energy per site $e(\chi)$ as a function of the bond dimension $\chi$. The energy per site only changes by about $3\times 10^{-6}$ as we increase the bond dimension from $203$ to $256$, suggesting that the error in the energy due to using a finite value of $\chi$ might be on that order of magnitude. In Fig. \ref{fig:Energy2} we then see that the energy converges to its extrapolated value $e(\infty) \approx -3.11229$ roughly as $\chi^{-2}$.

\subsection{Ground state entanglement}

From the TTN it is particularly simple to extract the spectrum of eigenvalues of reduced density matrices for particular blocks of spins, and thus compute the corresponding entanglement spectrum and entanglement entropy. 

Specifically, the upper bond index of the isometry $w^{(\tau,n)}$  corresponds to a block of $2^\tau$ sites of the 1D effective spin chain (or a rectangular block of $2^{\tau}\times L_y$ quantum spins of the initial 2D lattice model). The spectrum $\{p^{(\tau)}_{\alpha}\}_{\alpha=1}^{\chi}$ of eigenvalues of the reduced density matrix $\rho^{(\tau)}$ on that index can then be converted into the entanglement spectrum 
\begin{equation}
\left\{ \lambda_{\alpha}^{(\tau)} \equiv -\log_2 \left( p_{\alpha}^{(\tau)} \right)\right\}_{\alpha=1}^{\chi}
\end{equation}
and the entanglement entropy 
\begin{equation}
S^{(\tau)} \equiv -\sum_{\alpha=1}^{\chi} p_{\alpha}^{(\tau)} \log_2 \left( p_{\alpha}^{(\tau)} \right)
\end{equation}
for that block of spins.

\begin{figure}
  \includegraphics[width=8cm]{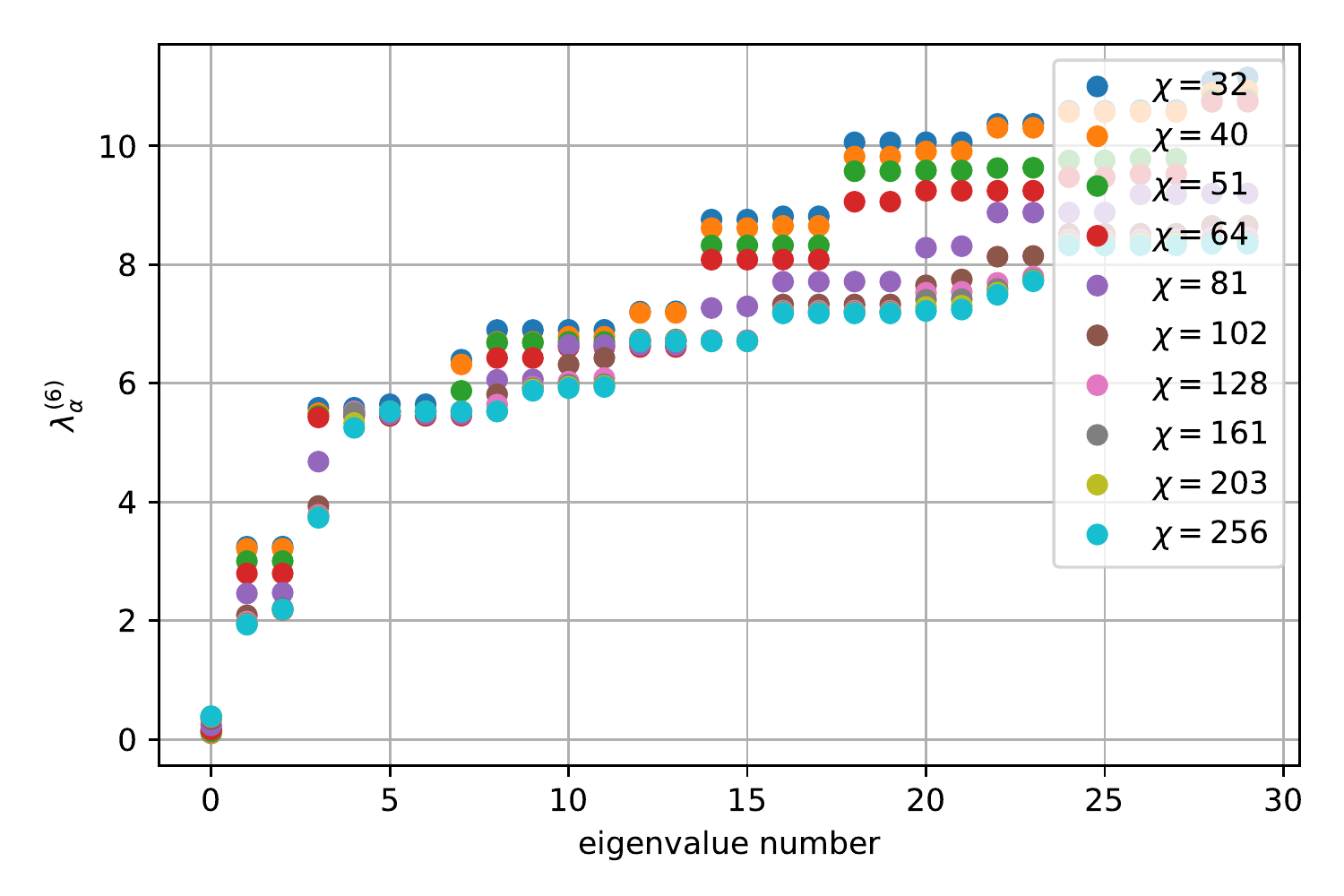}
\caption{
First 30 values of the entanglement spectrum $\{\lambda_{\alpha}^{(\tau)}\}_{\alpha=1}^{\chi}$ of the reduced density matrix assigned to an upper bond index of the 6th row (i.e. $\tau=6$) of isometries of the TTN, corresponding to a block of $2^{6}=64$ sites of the effective 1D spin chain. As expected, the lowest values converge faster than the larger ones with growing bond dimension $\chi$.
\label{fig:EntSpectrum} 
}
\end{figure}

\begin{figure}
  \includegraphics[width=8cm]{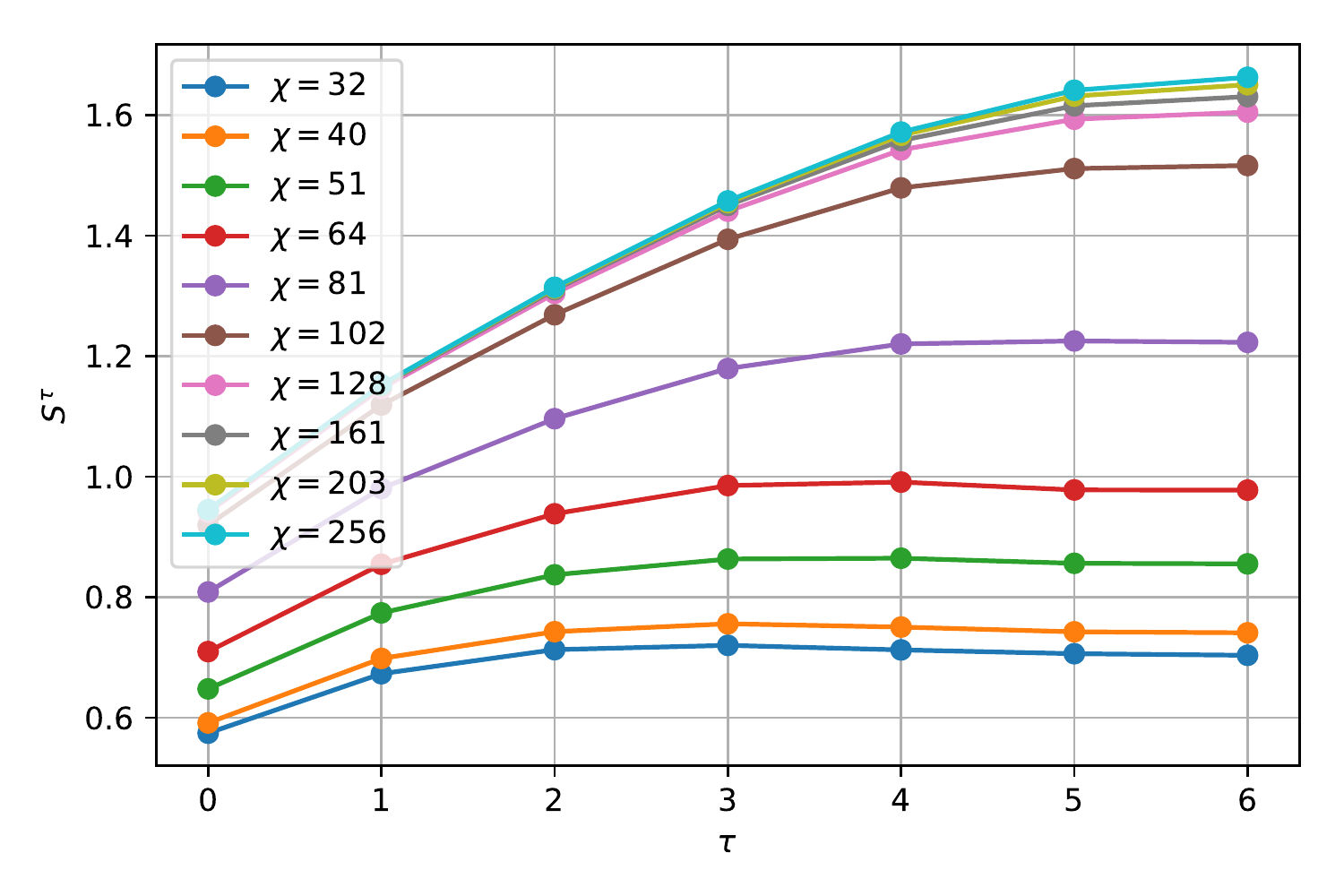}
\caption{
Entanglement entropy $S^{(\tau)}$ for a block of $2^{\tau}$ sites of the effective 1D quantum spin chain. As we increase the bond dimension $\chi$, the TTN variational ansatz is capable of better reproducing the entanglement structure of the ground state. We see that for the largest bond dimensions in the range $\chi=128-256$ the profile of entanglement entropies is already very stable. 
\label{fig:EntEntropy} 
}
\end{figure}

Fig. \ref{fig:EntSpectrum} shows the entanglement spectrum for $\tau=6$, that is for a block of $2^6 = 64$ sites of the effective 1D quantum spin chain or $2^6 \times 5 = 320$ sites of the 2D quantum Ising model on the thin torus. One can see that, as a function of the of the bond dimension $\chi$, the lower part of the spectrum converges faster than the upper part. Fig. \ref{fig:EntEntropy} then shows the scaling of entanglement entropy $S^{(\tau)}$ as a function of $\tau$, for different values of $\chi$. 
 
\begin{figure}
  \includegraphics[width=8cm]{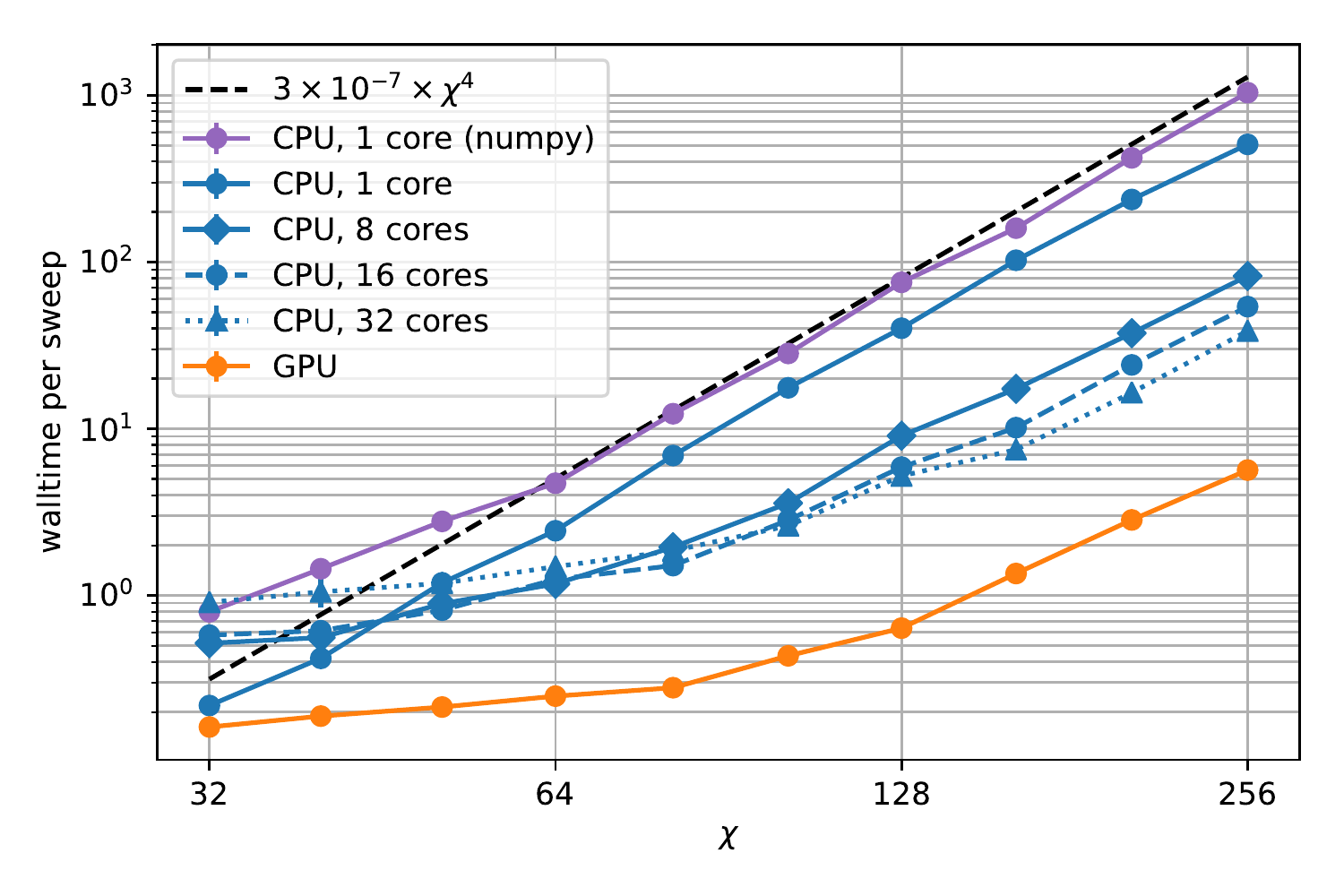}  %\includegraphics[width=8cm]{TTN_sweep_bench_tube_V100_v8.pdf}
  \caption{
Computational time as a function of the bond dimension $\chi$. For large $\chi$, the computational time on a single CPU (both numpy and TensorNetwork) scales as $O(\chi^4)$, as anticipated, with TensorNetwork being twice as fast as numpy for large $\chi$. On the GPU, the computational time does not yet reach the large $\chi$ scaling $O(\chi^4)$ but scales instead roughly as $O(\chi^3)$ for the largest bond dimensions we tested. Using clusters of $8,16,32$ CPUs reduces the gap with the GPU, although a GPU is still a factor $\times 6$ faster than $32$ CPUs.
\label{fig:cost} 
}
\end{figure}

\section{Computational time}

A highlight of TensorNetwork is that, thanks to running on top of TensorFlow, the same tensor network code \cite{download} can be used on different computational resources. We used TensorFlow v1.13.1 built with the Intel math kernel library (MKL). The computations described above were carried out using real numbers at 64 bit floating-point precision. We employed Google's cloud compute engine. For CPU computations we used Xeon Skylake with 1, 8, 16, and 32 cores. For GPU computations we used NVIDIA Tesla V100. For further reference, we also run equivalent numpy code using a single CPU.

The computational cost of the TTN algorithm scales as $\chi^4$ for sufficiently large $\chi$. This is indeed the scaling of both tensor-tensor contractions and of the SVD of the environment $E$ required to update an isometry $w$. There are also other steps, including permuting indices of rank-3 tensors, that scale as $\chi^3$.

Fig. \ref{fig:cost} shows the computational time required in order to update all the isometries in the TTN once (wall time per sweep). We see that for large bond dimension $\chi$, single CPU computations with code using either the numpy library or TensorNetwork both scale as $\chi^4$, as expected. However, using TensorNetwork is twice as fast. We also observe that for large bond dimension, using TensorNetwork with a GPU is about $100$ times faster than with a CPU. Moreover, with the range of tested bond dimension $\chi \leq 256$, the cost still scales roughly as $\chi^3$ on the GPU (larger values of $\chi$ will be tested in the near future). Finally, further optimizations are still required to fully take advantage of TPU architecture (work in progress), but early experiments suggest that the performance will likely exceed that of the GPU when those optimizations are completed.

\section{Conclusions}

This paper described a TTN algorithm for approximating the ground state of a quantum spin lattice model on a thin cylinder, implemented using TensorNetwork \cite{download}, an open source library that works on TensorFlow \cite{TensorFlow}. The code can be found here \cite{download}. We have used this sample code to find increasingly refined TTN approximations to the ground state of the transverse field Ising Hamiltonian on a periodic 2D lattice made of $L_x \times L_y = 128 \times 5 = 640$ quantum spins, with bond dimension $32 \leq \chi\leq 256$. Using TensorNetwork, we have seen that when running the code on a GPU, the computational time was reduced by a factor $\times 100$ compared to a single CPU. 

Code for other simulation algorithms for quantum systems based on tensor networks, such as MPS and MERA algorithms, will be similarly provided and discussed in subsequent papers.

\textbf{Acknowledgements.---} A. Milsted, M. Ganahl, and G. Vidal thank X for their hospitality. 
X is formerly known as Google[x] and is part of the Alphabet family of companies, which includes Google, Verily, Waymo, and others (www.x.company).
Research at Perimeter Institute is supported by the Government of
Canada through the Department of Innovation, Science
and Economic Development Canada and by the Province
of Ontario through the Ministry of Research, Innovation
and Science.

\section*{Appendix}

\begin{figure}
  \includegraphics[width=8cm]{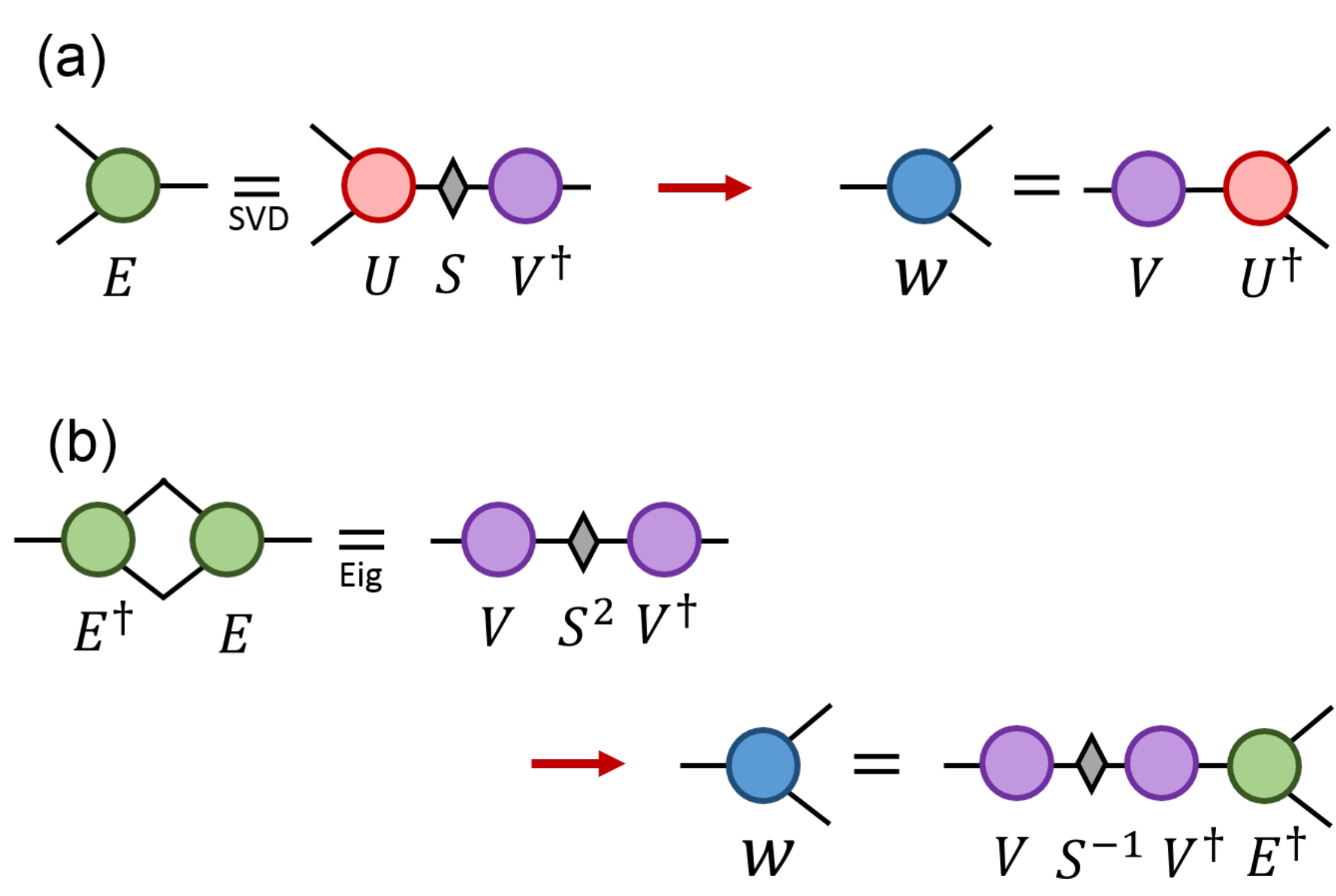}
\caption{
(a) Computation of an isometry $w$ from the corresponding environment tensor $E$, which can be regarded as a $\chi^2 \times \chi$ matrix. First the environment is decomposed in its singular value decomposition $E = USV^{\dagger}$, at cost $O(\chi^4)$. Then $w$ is built as $w = VU^{\dagger}$. 
(b) Alternative computation of of an isometry $w$ from the corresponding enviroment tensor $E$. This time first the squared environment $E^{\dagger}E$, which is a $\chi \times \chi$ matrix, is decomposed in its eigenvalue decomposition $E^{\dagger}E = V S^2 V^{\dagger}$. Then $w$ is built as $w = V S^{-1} V^{\dagger} E^{\dagger}$.  
\label{fig:AlternativeEnvironment} 
}
\end{figure}

Compared to a CPU, both GPU and TPU appear to provide very significant computational speed-ups on the order of $\times 100$-$1000$ for tensor-tensor contractions involving large tensors, but more modest speed-ups for matrix factorizations such as a singular value decomposition (SVD) or eigenvalue decomposition (EVD). In those tensor network algorithms, such as MERA algorithms, where the cost of the required tensor-tensor multiplications and SVD scale e.g. as $O(\chi^9)$ and $O(\chi^6)$ respectively, the use of GPUs and TPUs is expected to lead to massive savings in computational time. However, in a TTN where tensor-tensor multiplications and SVD scale both as $O(\chi^4)$, GPUs and TPUs will lead to less spectacular gains.

In our current TTN algorithm, it is possible to replace the $O(\chi^4)$ SVDs with $O(\chi^4)$ tensor-tensor multiplications and $O(\chi^3)$ EVDs, see Fig. \ref{fig:AlternativeEnvironment}. In this way, larger speed-ups than the ones reported in the main text are expected. However, the squaring of the environment $E$ in Fig. \ref{fig:AlternativeEnvironment}(b) leads to a loss of half of the numerical precision. In those simulations where this is not a problem (e.g. because the error due to a finite bond dimension is more important than the loss of numerical precision due to squaring the environment), it might then convenient to use an $O(\chi^3)$ EVD instead of an $O(\chi^4)$ SVD.


\begin{thebibliography}{99}
\bibitem{Comparison} The TTN is also between MPS and MERA when it comes to the scaling of the computational cost with the bond dimension $\chi$ of the tensor. In 1D it is $O(\chi^3)$ for MPS, $O(chi^4)$ for the TTN, and $O(\chi^7)$ for the MERA, although the required bond dimension $\chi$ to achieve a given accuracy (e.g. in ground state energy) depends on the tensor network and therefore the above scaling is not representative of the relative performance of these tensor networks, see e.g. Fig. 12 in \cite{criticalMERA}.
\end{thebibliography}
\end{document}